\documentclass{appolb}
\usepackage{epsfig}

\begin{document}
\title{Mach cone shock waves at RHIC%
\thanks{Presented at the XXXVII International symposium on Multiparticle Dynamics}%
}

\author{
J\"org Ruppert\footnote{presenting author}
\address{Department of Physics, McGill University, 3600 University Street, Montreal, QC, H3A 2T8 Canada}
\and
Thorsten Renk
\address{Department of Physics, PO box 35 FIN-40014, University of Jyv\"askyl\"a, Finland, and 
Helsinki Institute of Physics, PO Box 64, FIN-00014, University of Helsinki, Finland.}
}

\maketitle
\begin{abstract}
Energy and momentum lost by hard jets propagating through hot and dense nuclear matter have to be redistributed in the medium. It has been conjectured that collective sound modes
are excited. Those lead to Mach cone nuclear shock waves in the nuclear medium 
that are shown to account for three and four particle angular correlation structures of hadrons with a (semi-)hard trigger hadron in heavy-ion collisions at RHIC. 
\end{abstract}
  
\section{Introduction}
The quenching of QCD jets created in relativistic nuclear collisions has been proposed as an important indicator of the creation of a quark-gluon plasma \cite{Jet1}. It is extensively studied theoretically  and experimentally at RHIC. The main emphasis in most theoretical studies (before 2005) has been solely on the description of the radiative energy loss which the leading parton suffers while
traversing the nuclear medium due to the emission of partonic secondaries. In this paper we focus on another aspect of the in-medium jet physics, namely the question if the jets traversing the medium can transfer energy and momentum to collective modes in the nuclear medium that might be able to account for the emergence of peculiar signals in the particle correlation measurements.
Recently, measurements of two- and three-particle correlations involving one hard trigger particle have shown a surprising splitting of the away side peak for all centralities but peripheral collisions, qualitatively very different from a broadened away side peak observed in p-p or d-Au collisions \cite{PHENIX-2pc}. Interpretations in terms of colorless \cite{Stoecker,Shuryak} and colored \cite{Ruppert1} sound modes have been suggested for an explanation of this phenomenon.  For an overview also discussing alternative mechanisms, see \cite{Ruppert2} and references therein.
While the microscopic excitation mechanism of the colorless modes is under investigation \cite{Neufeld}, a detailed theoretical study of the experimentally observable correlations is already now possible if this mechanism is assumed to effectively excite the mode.
In the following, we discuss how such shockwaves lead to observable correlation signals in the dynamical environment of a heavy-ion collision. 

\section{The model}
A Monte Carlo simulation of the hard back to back process in the medium is performed. There are four main stages in the modeling: 1) the primary hard pQCD process, 2) the description of the soft medium, 3) the energy loss from hard to soft degrees of freedom, 4) the simulation of the shockwave propagation and its modification of the soft medium until decoupling. For brevity we only present a  sketch of the model here, a detailed description of the model can be found in \cite{Mach1}.
The soft medium is described by a parametrized evolution model \cite{Parametrized} which gives a good description of the bulk matter transverse 
momentum spectra and HBT correlation radii. The energy loss for a given parton path inside this medium is described probabilistically by $P(\Delta E,E)_{\rm path}$, which is the probability for a hard parton of energy $E$ to lose energy $\Delta E$ while traversing the medium in the ASW formalism \cite{JetScaling}.  
Since we are interested in the energy deposition on average in a given volume we focus on the distribution of the average energy $<\Delta E>=\int_0^\infty P(\Delta(E)) \Delta E d\Delta E$ along the paths to infer $dE/dx$, see Fig. 1 in \cite{Mach1}. We assume that a fraction f of the energy lost to the medium excites a shockwave characterized by a dispersion relation $E=c_s p$ where $c_s$ is the speed of sound inferred from the equation of state by $c_s^2=\partial p / \partial \epsilon$.
The dispersion relation determines the initial angle of propagation of the shock front  as $\phi= \arccos c_s$. 
The time dependent energy transfer to the mode per unit time is given by $f dE/d\tau$.
Each piece of the front is propagated with the local speed of sound through the medium. Once a wavefront has evolved until it satisfied freeze-out condition $T=T_F$ it cannot propagate further leading to an additional boost at freeze-out. The Cooper-Frye formalism is employed to calculate the hadronic distribution accordingly. On the near-side the trigger condition is realized. The important role of longitudinal and transverse expansion as well as flow effects on the observed correlation signal are discussed in \cite{Mach1}.

\section{Results}

\begin{figure}[htb]
\begin{center}
\epsfig{file=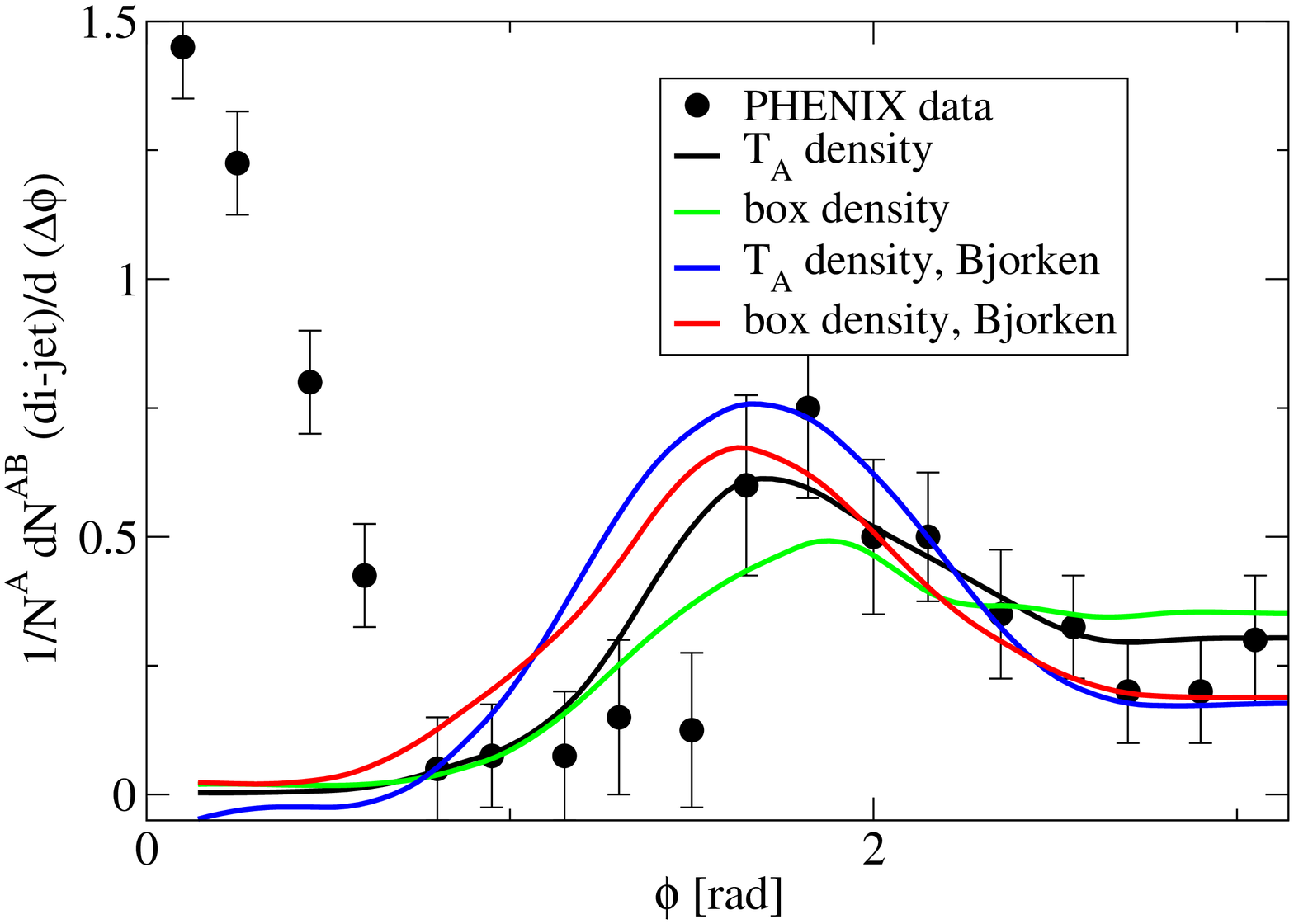, height=4.5cm}
\end{center}
\caption{\label{F-1}Correlation signal on the away side for different soft background medium density distributions and evolutions (see text, not acceptance-averaged).}
\end{figure}

\begin{figure}[htb]
\epsfig{file=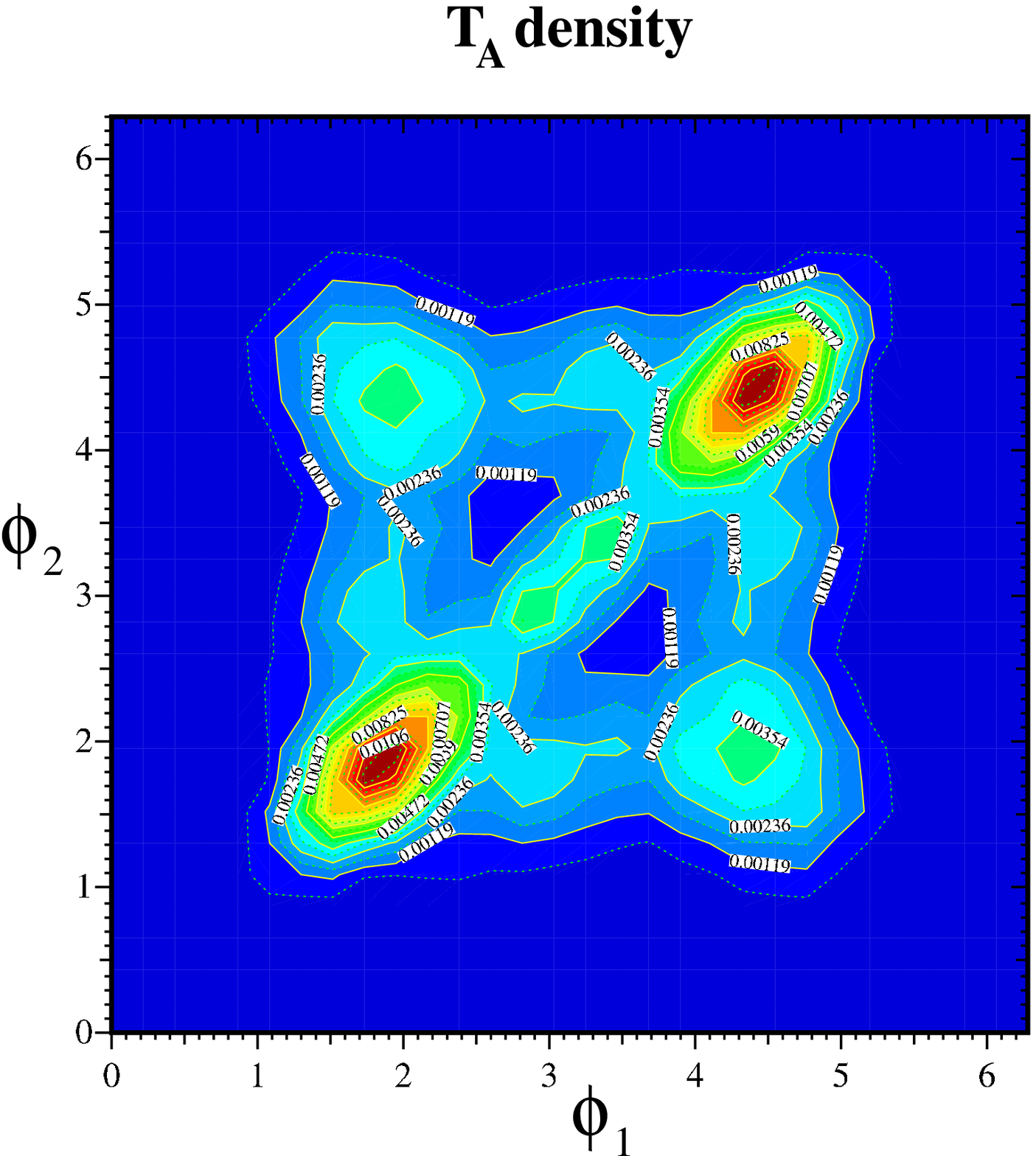, width=5.9cm} \epsfig{file=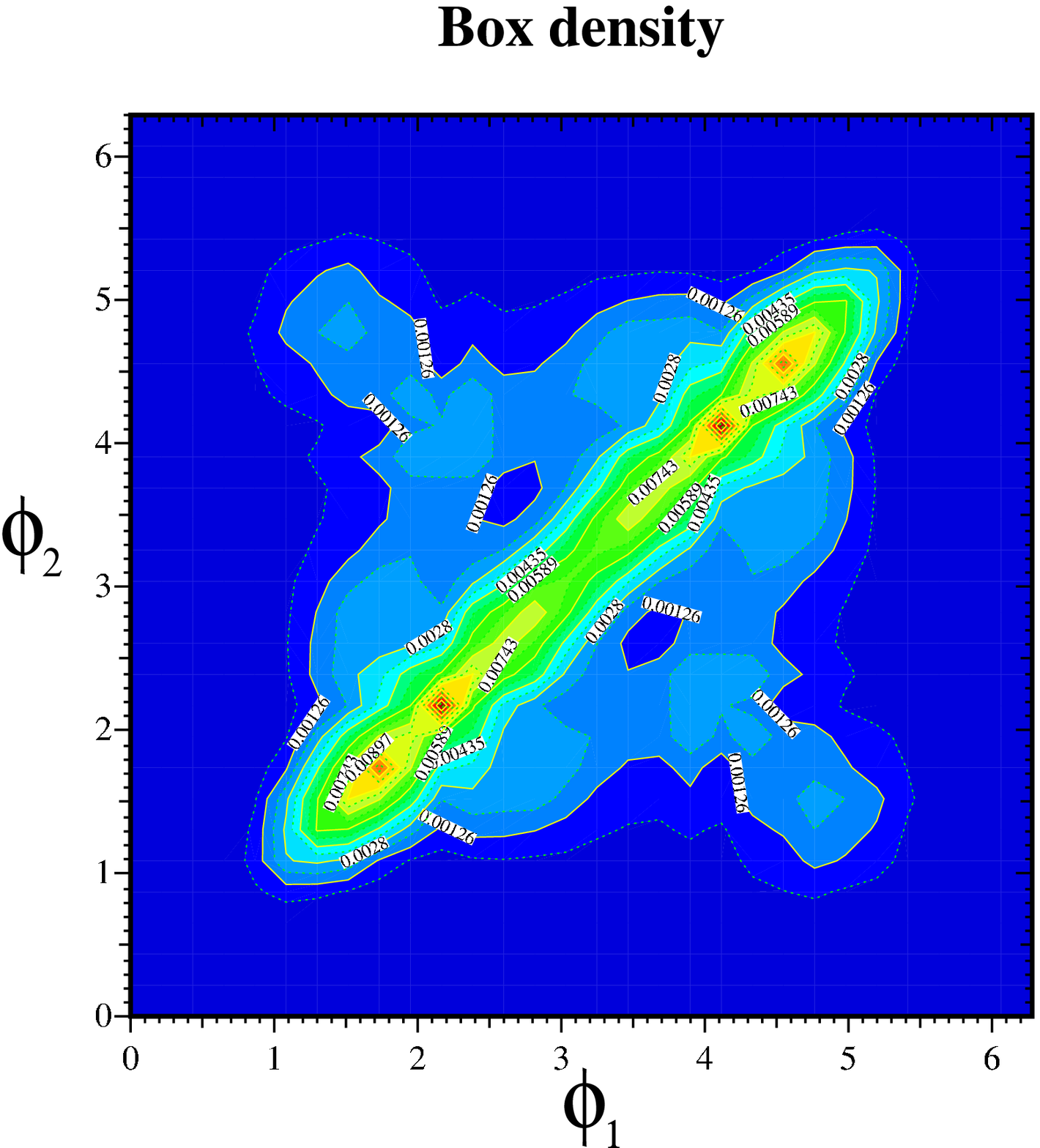, width=5.9cm}
\vspace*{-0.30cm}
\caption{\label{F-2} 3-particle correlation strength for associate hadrons with $1.0 < p_T < 2.5$ GeV as a function of angles $\phi_1$ and $\phi_2$ with the trigger for a nuclear profile function shaped and a box shaped medium density distribution.}
\end{figure}

In order to show how the excitation of a collective mode can account for the observed correlation signal in a quantiative manner, we show in Fig. \ref{F-1}  the correlation signal and its sensitivity to different descriptions of the soft background. We chose an energy independent $f=0.75$ which accounts for the observed signal.
We show results for two different expansion patterns (Bjorken and non-Bjorken evolution) and two different transverse densities (box and nuclear profile $T_A$).
Due to the position of the freeze-out hypersurface at large radii for the box density, the shockwave gets on average longer exposure to transverse flow in this scenario which leads to a less pronounced maximum since transverse flow can erase a peak if flow and shockwave are not aligned. The angle is larger for longitudinal Bjorken expansion which is due to the fact that initial cooling is rapid and therefore the averag temperature quickly approaches the phase transition temperature $T_c$ where $c_s$ is small.
In this specific calculation we suppressed a full rapidity averaging, leading to a somewhat larger angle than expected if averaging were performed. 
In Fig. \ref{F-2} we show the resulting 3- particle correlation for non-Bjorken evolution with initial box density and nuclear density profile, for details of the calculation see \cite{Mach1}. 
The region close to  the trigger is not studied. Our results fit nicely with the fact that transverse flow can erase in some configurations 
one or even both of the wings of the cone. In this case, only the diagonal region is populated whereas
if both wings appear the off diagonal maxima are populated. In this sense 2- and especially 3-particle correlation measurements do not only reveal information about the smallness of $c_s$ in the evolution, indicating the existence of a (cross-over) phase transition to a quark gluon plasma, but
also can eventually be used to further constrain our understanding of the evolution of the soft-matter at RHIC. 
 This work was supported by the Natural Science and Engineering  Research Counsil of Canada and
the Academy of Finland. 


\vspace*{-0.35cm}
\begin{thebibliography}{00}
\bibitem{Jet1}
  M.~Gyulassy and X.~N.~Wang,
  Nucl.\ Phys.\ B {\bf 420}, 583 (1994),
  T.~Renk, J.~Ruppert, C.~Nonaka and S.~A.~Bass,
  Phys.\ Rev.\  C {\bf 75} (2007) 031902,
 G.~Y.~Qin et al.,  arXiv:0710.0605 [hep-ph].    
\bibitem{PHENIX-2pc}
S.~S.~Adler et al. [PHENIX collaboration].
nucl-ex/0507004.
\bibitem{Stoecker}
  H.~St\"ocker,
  Nucl.\ Phys.\ A {\bf 750} (2005) 121.
\bibitem{Shuryak}
  J.~Casalderrey-Solana, E.~V.~Shuryak and D.~Teaney,
  J.\ Phys.\ Conf.\ Ser.\  {\bf 27}, 22 (2005), hep-ph/0602183.
 \bibitem{Ruppert1}
  J.~Ruppert and B.~M\"uller,
  Phys.\ Lett.\ B {\bf 618} (2005) 123;
   J.~Ruppert,
   J.\ Phys.\ Conf.\ Ser.\  {\bf 27} (2005) 217,
   B.~M\"uller and J.~Ruppert,
   nucl-th/0507043,
    J.~Ruppert,
   hep-ph/0510386.
\bibitem{Ruppert2}
 J.~Ruppert,
   Nucl.\ Phys.\ A {\bf 774} (2006) 397. 
\bibitem{Neufeld}
 B. M\"uller, B. Neufeld, J. Ruppert, in preparation
\bibitem{Mach1}
  T.~Renk and J.~Ruppert,
  Phys.\ Rev.\ C {\bf 73} (2006) 011901, 
  Phys.\ Rev.\  C {\bf 76}, 014908 (2007),
   hep-ph/0701154.
\bibitem{Parametrized}
  T.~Renk,
  Phys.\ Rev.\ C {\bf 70} 021903 (2004).
 \bibitem{JetScaling}
  C.~A.~Salgado and U.~A.~Wiedemann,
  Phys.\ Rev.\ Lett.\  {\bf 89}, (2002) 092303, 
  Phys.\ Rev.\ D {\bf 68}, (2003) 014008.
   
\end{thebibliography}
\end{document}